\documentclass[APS,twocolumn,reprint,amsmath,amssymb]{revtex4}
\usepackage{CJK}
\usepackage{indentfirst}
\usepackage{float}
\usepackage{graphicx}
\usepackage{dcolumn}
\usepackage{bm}
\usepackage[colorlinks=true, linkcolor=blue, citecolor=blue, urlcolor=blue, allcolors=blue]{hyperref}
\DeclareGraphicsExtensions{.eps,.ps,.bmp,.jpg,.pdf,.png}

\begin{document}

\title{Gate-Switchable Josephson Diode Effect in Graphene via Magnetochiral Anisotropy}
\author{Chuan-Shuai Huang\textsuperscript{1}}
\email[Email: ]{245748852@qq.com}

\author{Xing He\textsuperscript{2}}
\author{Ruihao Zhang\textsuperscript{1}}

\email[Email: ]{ruihao.zhang@sru.edu.cn}

\affiliation{\textsuperscript{1}School of Optoelectronic Information Engineering, Shangrao Normal University, Shangrao 334001, China}
\affiliation{\textsuperscript{2}School of Mathematics and Physics, Shangrao Normal University, Shangrao 334001, China}

\date{\today}

\begin{abstract}
The Josephson diode effect originating from magnetochiral anisotropy has thus far been realized primarily in conventional two-dimensional electron gases. Here, we propose a graphene-based Josephson junction as a broadly gate-tunable platform for nonreciprocal supercurrent transport. We demonstrate that the nonreciprocal critical current not only reverses its sign when graphene is switched from $n$- to $p$-type doping via electrostatic gating, but its efficiency can also be continuously tuned from zero up to $\sim 40\%$. This gate-controlled polarity switching and efficiency modulation---unique to the chiral nature of Dirac fermions in graphene---has no counterpart in conventional semiconductor systems, positioning the graphene Josephson diode as a highly versatile component for advanced superconducting circuits.
\end{abstract}

\maketitle

\section{Introduction}

Nonreciprocal transport, where $R(+I)\neq R(-I)$, arises from simultaneous breaking of time-reversal and inversion symmetries. In noncentrosymmetric conductors, this yields magnetochiral anisotropy (MCA), a nonlinear response described by $R=R_0[1+\gamma\,\hat{\mathbf{e}}_z\cdot(\mathbf{B}\times\mathbf{I})]$. Here $\gamma$ is the rectification coefficient proportional to the Rashba spin--orbit coupling (RSOC) strength, $\mathbf{B}$ is the applied magnetic field breaking time-reversal symmetry, $\mathbf{I}$ is the current direction, and $\hat{\mathbf{e}}_z$ denotes the axis along which inversion symmetry is broken; these three vectors must be mutually orthogonal to maximize the effect \cite{Tokura2018,He2018,Wang2022,Ideue2017}.

In superconductors, MCA manifests as a superconducting diode effect (SDE), where the critical current depends on its direction, allowing dissipationless transport in one direction only \cite{Ando2020,Nadeem2023,Ren2026,Huang2026}. Such a diode effect enables novel superconducting circuit elements with built-in current rectification, holding promise for energy-efficient superconducting spintronics and quantum computing \cite{Ma2025,Dirnegger2026}.
In Josephson junctions, the analogue is known as the Josephson diode effect (JDE) \cite{Hu2007,Yokoyama2014,Zhang2022,Davydova2022,Schulz2025,Mahmud2025,Nikolic2026}, which manifests as a nonreciprocal current--phase relation (CPR): $I(\phi) \neq I(-\phi)$. Microscopically, in the short-junction limit, the supercurrent is carried by Andreev bound states (ABSs) \cite{Andreev1966,Beenakker1991,Golubov2004,Costa2018b,Park2024}, whose energy spectrum acquires an asymmetric phase shift under the combined action of RSOC and an in-plane Zeeman field \cite{Szombati2016,Mayer2020,Strambini2020,Costa2023theory,Djurdjevic2025,Qin2026}. A sizable diode effect further requires a skewed CPR with substantial higher harmonics \cite{Tanaka2022,Baumgartner2022a}.

Experimentally, the JDE was first realized in Al/InAs/Al junctions \cite{Baumgartner2022}, and to date, all MCA-based Josephson diodes have been fabricated on conventional semiconductor 2DEGs \cite{Baumgartner2022,Baumgartner2022a,Costa2023,Reinhardt2023,Shin2026}. These platforms have enabled systematic studies of the interplay between SOC, Zeeman fields, and supercurrent nonreciprocity, revealing the MCA mechanism, $0$-$\pi$-like transitions \cite{Costa2023}, and the link between the diode effect and the anomalous Josephson effect \cite{Reinhardt2023}. Gate-tunable polarity switching has indeed been demonstrated in 2DEG-based devices \cite{Shin2026}, but the underlying mechanism relies on modifying the Rashba SOC strength or carrier density. A fundamentally different route to polarity control becomes available in graphene, where switching between electron and hole doping reverses the sign of the Zeeman-induced Fermi-surface shift \cite{Inglot2014}, and is thus expected to yield an opposite MCA.

\begin{figure}[t]
\centering
\includegraphics[width=7cm]{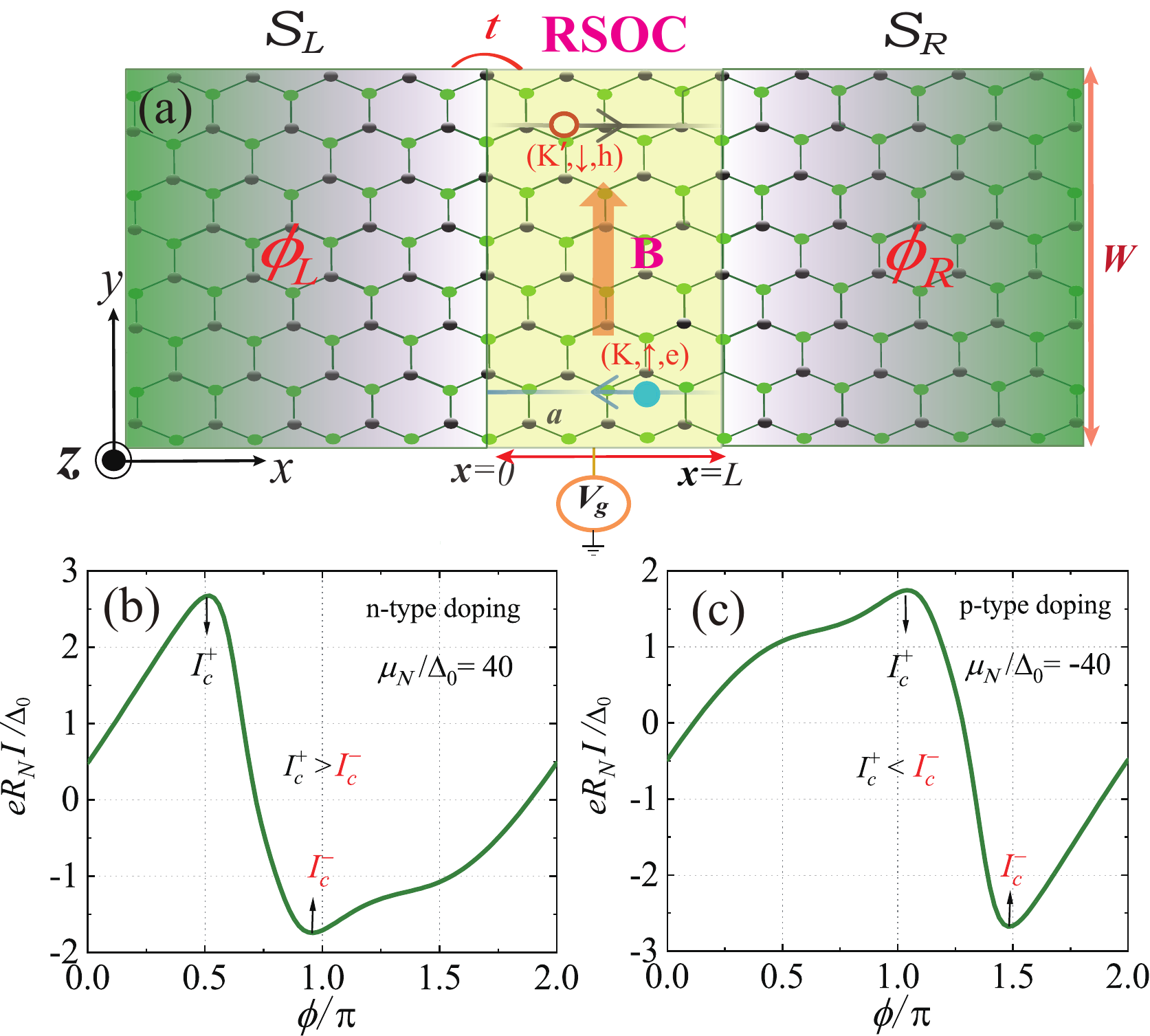}
\caption{(a) Schematic of a short SNS Josephson junction based on monolayer graphene. The central N region ($0<x<L$) is subject to a proximity-induced in-plane Zeeman field $m_y$ and Rashba spin--orbit coupling $t_R$. A gate voltage $V_g$ controls the carrier density and type. (b), (c) Calculated current--phase relations for $n$-type ($\mu_N = +40\Delta_0$) and $p$-type ($\mu_N = -40\Delta_0$) doping, respectively. Parameters: $m_y = t_R = 10\Delta_0$, $L=100$, $T/T_c=0.1$. $R_N$ is the Sharvin resistance of a perfectly transparent graphene junction in the normal state.}
\label{fig1}
\end{figure}

Motivated by this unique feature, we propose a single-layer graphene-based Josephson junction as a highly versatile JDE platform. Although graphene intrinsically possesses weak spin-orbit coupling, sizable RSOC and exchange fields can be induced via proximity to adjacent materials \cite{Wei2016,Peralta2019,Tseng2022,Dai2026,Varykhalov2008,Marchenko2012,Avsar2014,Gmitra2015,WangZ2015,Rakitskii2026}. As illustrated in Fig.~\ref{fig1}(a), the device consists of a graphene strip contacted by two superconducting electrodes, with a gate electrode controlling the doping level $\mu_N$ in the normal region. Applying an in-plane magnetic field and proximity-induced RSOC breaks both time-reversal and inversion symmetries, enabling MCA.

The calculated CPRs for $n$- and $p$-type doping are shown in Figs.~\ref{fig1}(b) and (c). Strikingly, the nonreciprocal critical currents reverse sign when switching carrier type, demonstrating gate-controlled diode polarity. Moreover, the diode efficiency can be tuned continuously from zero up to $\sim$40\% by varying the gate voltage---a giant tunability rooted in graphene's chiral band structure that far exceeds that of conventional 2DEG diodes. Additionally, $0$-$\pi$-like transitions in the CPR at larger Zeeman fields are predicted, consistent with recent experiments \cite{Costa2023}. Crucially, this MCA-based mechanism is fundamentally distinct from other graphene JDE proposals \cite{Wei2022,DiezMerida2023,Hu2023,ValleyPolarizedMATBG,Ilic2024,Bhardwaj2025,Rothstein2026,Pezzini2026}. The results establish graphene as a powerful platform for gate-programmable Josephson diodes, combining strong proximity-induced spin-orbit and exchange fields with the unique ability to switch carrier types electrostatically.

\section{Model and Formalism}

We model the central N region of the graphene junction using a nearest-neighbor tight-binding Hamiltonian that includes RSOC and an in-plane Zeeman field \cite{Kane2005,Qiao2010}:
\begin{equation}
\begin{aligned}
H = &-t\sum_{\langle i,j\rangle\alpha} c_{i\alpha}^\dagger c_{j\alpha} - \mu_i \sum_{i\alpha} c_{i\alpha}^\dagger c_{i\alpha}
- m_y \sum_{i\alpha\beta} c_{i\alpha}^\dagger \sigma_y^{\alpha\beta} c_{i\beta} \\
&+ i t_R \sum_{\langle i,j\rangle\alpha\beta} c_{i\alpha}^\dagger ( \bm\sigma \times \hat{\bm d}_{ij} )^z_{\alpha\beta} c_{j\beta},
\end{aligned}
\end{equation}
where $c_{i\alpha}^\dagger$ ($c_{i\alpha}$) creates (annihilates) an electron with spin $\alpha$ on site $i$, $\bm\sigma$ is the vector of Pauli matrices, and $\langle i,j\rangle$ denotes nearest neighbors. The first term is the nearest-neighbor hopping ($t=1$ eV). The second term is the chemical potential; we assume a large, fixed $\mu_S$ in the superconducting leads, while $\mu_N$ in the N region is tunable by a gate voltage $V_g$ \cite{CastroNeto2009,Beenakker2008,Kim2013}. The third term is the Zeeman coupling due to an in-plane magnetic field $\mathbf{B} = B \hat{\bm e}_y$. Although graphene's intrinsic $g$-factor is small, a significant exchange field $m_y$ can be induced via proximity to a magnetic insulator (e.g., EuS) under an applied field \cite{Wei2016,Peralta2019,Tseng2022,Dai2026}. The fourth term represents the extrinsic RSOC with strength $t_R$, which can reach $\sim$10 meV via proximity to transition-metal dichalcogenides or intercalated layers \cite{Varykhalov2008,Marchenko2012,Avsar2014,Gmitra2015,WangZ2015,Rakitskii2026}.

In the superconducting regions, we introduce an on-site $s$-wave pair potential $V_{\text{pair}}^{L(R)} = \sum_i \Delta e^{i\phi_{L(R)}} (c_{i\uparrow}^\dagger c_{i\downarrow}^\dagger + \text{H.c.})$ via proximity to external superconducting contacts \cite{Nanda2017,Lee2018}, but omit RSOC and Zeeman terms there. The temperature-dependent BCS gap is $\Delta(T) = \Delta_0 \tanh(1.74\sqrt{T/T_c - 1})$ with $\Delta_0 = 1.76 k_B T_c$. We set $\Delta_0 = 0.001t$ as our energy unit.

The supercurrent through the $l$th column in the N region is computed using the recursive Green's function technique \cite{Zhang2016,Minutillo2021,Liang2008,Singh2025}:
\begin{equation}
\begin{aligned}
I = \frac{1}{2\pi\hbar} \sum_{k_y} \int_{-\infty}^{+\infty} dE &\text{Tr}\bigl[ H_{l,l+1} \check e G^{<}_{l+1,l}(E,k_y) \\
&- \check e H_{l+1,l} G^{<}_{l,l+1}(E,k_y) \bigr],
\end{aligned}
\end{equation}
where $k_y = 2\pi n/W$ is the transverse momentum ($W \gg L$ ensures quantization) \cite{footnote}, $\check e = -e \tau_3 \otimes \sigma_0$ is the charge matrix, and $H_{l,l+1}$ couples layers $l$ and $l+1$. The lesser Green's function in equilibrium is calculated as $G^< = (G^a - G^r) f(E)$ with $f(E)$ the Fermi--Dirac distribution. The retarded Green's function is $G^r = [E\hat I - H_N - \Sigma_L^r - \Sigma_R^r]^{-1}$, where the self-energies $\Sigma_{L,R}^r$ account for coupling to the superconducting leads and are obtained recursively \cite{Caroli1971,LopezSancho1984,Datta1997}. The ABS spectra are extracted from the local density of states at column $l$ \cite{Fu2022}:
\begin{equation}
\rho_l = -\frac{1}{\pi} \text{Im}\{\text{Tr}[G^r(l,l)]\},
\end{equation}
with peaks inside the superconducting gap corresponding to ABS energies at a given phase difference $\phi = \phi_R - \phi_L$.

\begin{figure}[t]
\centering
\includegraphics[width=8cm]{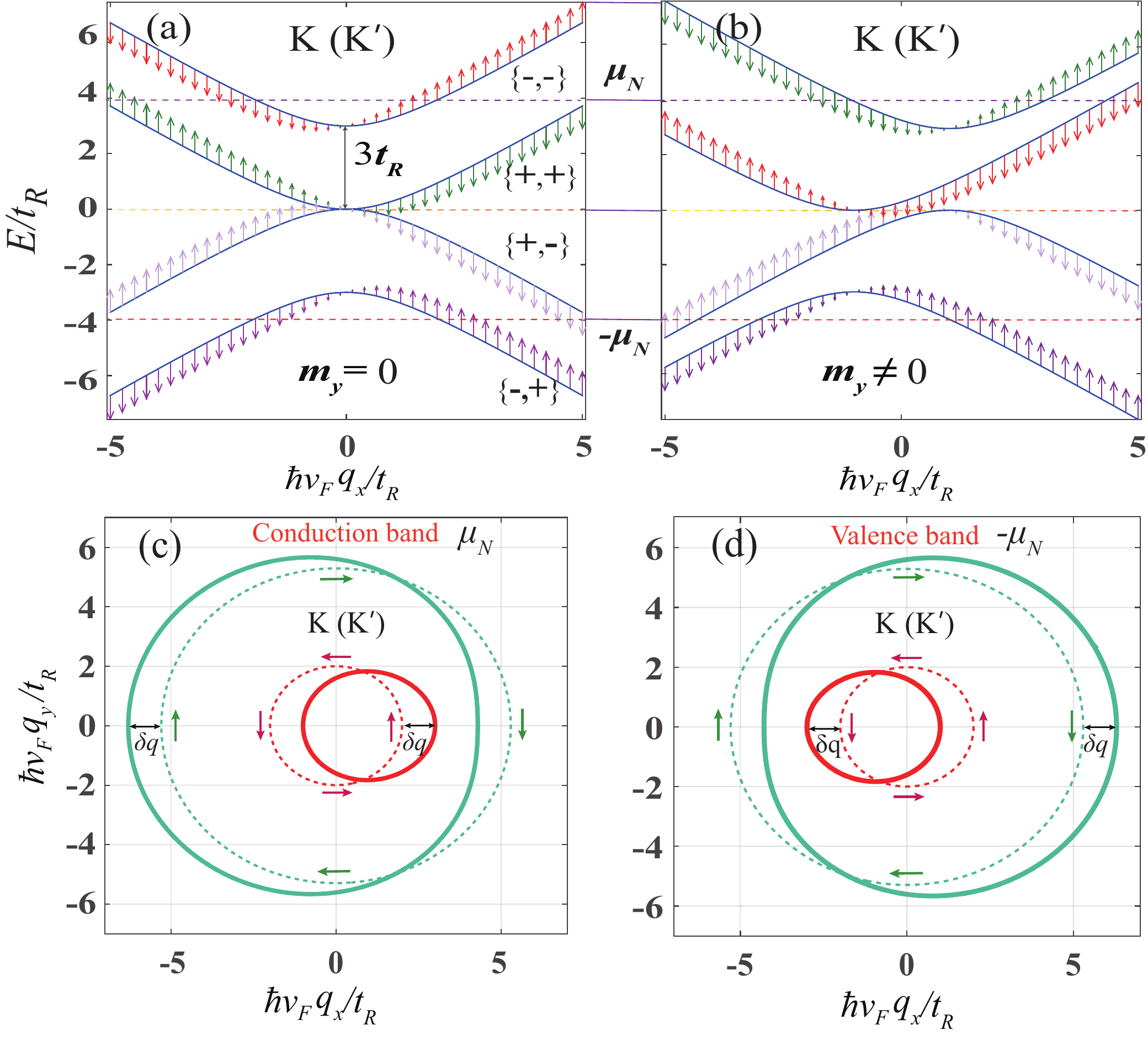}
\caption{Band structure and Fermi contours for graphene with Rashba spin-orbit coupling and Zeeman field. (a) Low-energy band structure near Dirac points for $m_y=0$, showing the four Rashba-split bands. Arrows indicate the in-plane spin polarization. (b) Same as (a) but with finite Zeeman field $m_y=t_R$, which lifts spin degeneracy and induces energy offsets. (c) Schematic spin-textured Fermi contours for the conduction band ($\mu\nu=++$ or $--$, depending on doping) with $\mu=+1$ (outer contour) and $\mu=-1$ (inner contour). (d) Fermi contours for the valence band ($\mu\nu=+-$ or $-+$). The spin polarizations are locked perpendicular to the momentum (black arrows) and are identical in K and K$'$ valleys. The Zeeman field induces opposite momentum shifts $\delta q_{\mu\nu}$ along $x$ (red arrows) for the two spin-split Fermi surfaces within each band.}
\label{fig2}
\end{figure}

\section{Band Structure and Andreev Bound States}
\label{sec:band}

\subsection{Band structure}

We focus on low-energy excitations near the Dirac points. For $m_y/t_R \rightarrow 0$, the spectrum is approximately
\begin{equation}
E_{\mu\nu} = \mu\nu \sqrt{(\hbar v_F q)^2 + \lambda_R^2} - \mu \lambda_R,
\end{equation}
where $q = |\bm q|$ is measured from the K (K$'$) point, $\lambda_R = 3t_R/2$, $\mu,\nu = \pm$ are band indices, and $v_F = 3 a t/2 \approx 1\times10^6$ m/s. Fig.~\ref{fig2}(a) shows the four bands for $m_y=0$, revealing the characteristic Rashba splitting with spin-momentum locking. The in-plane spin polarization in the state $| \psi_{\mu\nu} \rangle$ for $m_y=0$ is given by \cite{Rashba2009b}
\begin{equation}
\bm S_{\mu\nu}(\bm q) \equiv \langle \psi_{\mu\nu} | \bm\sigma | \psi_{\mu\nu} \rangle = \frac{\mu \hbar v_F (\bm q \times \hat{\bm z})}{\sqrt{\lambda_R^2 + (\hbar v_F q)^2}},
\end{equation}
which is locked perpendicular to the momentum, as indicated by the arrows in Fig.~\ref{fig2}(a).

Adding a Zeeman field $m_y$ breaks time-reversal symmetry, shifting the bands by $\langle \psi_{\mu\nu} | m_y \sigma_y | \psi_{\mu\nu} \rangle$ [Fig.~\ref{fig2}(b)]. Crucially, this Zeeman shift induces a momentum offset $\delta q_{\mu\nu}$ along the transport direction for each spin-split Fermi surface. Within first-order perturbation theory, the energy change $\Delta E_{\mu\nu} = \langle \psi_{\mu\nu} | m_y \sigma_y | \psi_{\mu\nu} \rangle$ must be compensated by a shift in momentum $\delta q_{\mu\nu}$ along the group velocity direction to maintain the Fermi energy. This leads to the approximate relation
\begin{equation}
\delta q_{\mu\nu} \simeq - \frac{\langle \psi_{\mu\nu} | m_y \sigma_y | \psi_{\mu\nu} \rangle_{\text{FS}}}{\bm v_{\mu\nu} \cdot \hat{\bm x}},
\label{eq:momentum_shift}
\end{equation}
where $\bm v_{\mu\nu} = \partial E_{\mu\nu}/\hbar \partial \bm q$ is the group velocity, and $\langle \cdots \rangle_{\text{FS}}$ denotes averaging over the Fermi surface. The minus sign indicates that the momentum shift opposes the energy change. Because $\bm S_{\mu\nu}$ has opposite signs for the two spin-split bands, their momentum shifts are also opposite, as illustrated by the black arrows in Figs.~\ref{fig2}(c) and (d).

\subsection{ABSs at zero transverse momentum}
\label{sec:ABS_k0}

In a short SNS junction, the supercurrent is carried by discrete ABSs localized in the central normal region, formed by the closed-loop motion of electrons and holes undergoing Andreev reflections at the two superconductor-normal interfaces. These subgap states arise from the phase accumulation during each round trip, and their energies depend sensitively on the superconducting phase difference $\phi=\phi_R-\phi_L$ across the junction.

As discussed previously, the Zeeman-induced momentum shift $\delta q_{\mu\nu}$ has profound consequences for these ABSs. During Andreev reflection, electrons and holes acquire opposite signs of $\delta q_{\mu\nu}$, resulting in a net phase accumulation $\delta\phi_{\mu\nu} = 2\delta q_{\mu\nu} L$ per round trip across the junction of length $L$. In the short-junction limit ($L \ll \xi$), the ABS energies at $k_y=0$ are approximately \cite{Beenakker1992}
\begin{equation}
\varepsilon_{\mu\nu} = \pm \Delta \sqrt{1 - \tau_{\mu\nu} \sin^2(\phi/2 + \delta\phi_{\mu\nu}/2)},
\label{eq:ABS_energy}
\end{equation}
where $\tau_{\mu\nu}$ is the normal-state transmission probability for channel $(\mu,\nu)$. The CPR follows from the phase derivative of the ABS energies \cite{Linder2008,BlackSchaffer2010}:
\begin{equation}
I(\phi) \simeq -\frac{2e}{\hbar} \sum_{\mu\nu} \frac{\partial \varepsilon_{\mu\nu}}{\partial \phi} \tanh(\varepsilon_{\mu\nu}/2k_B T).
\label{eq:CPR_from_ABS}
\end{equation}

\begin{figure}
\centering
\includegraphics[width=\linewidth]{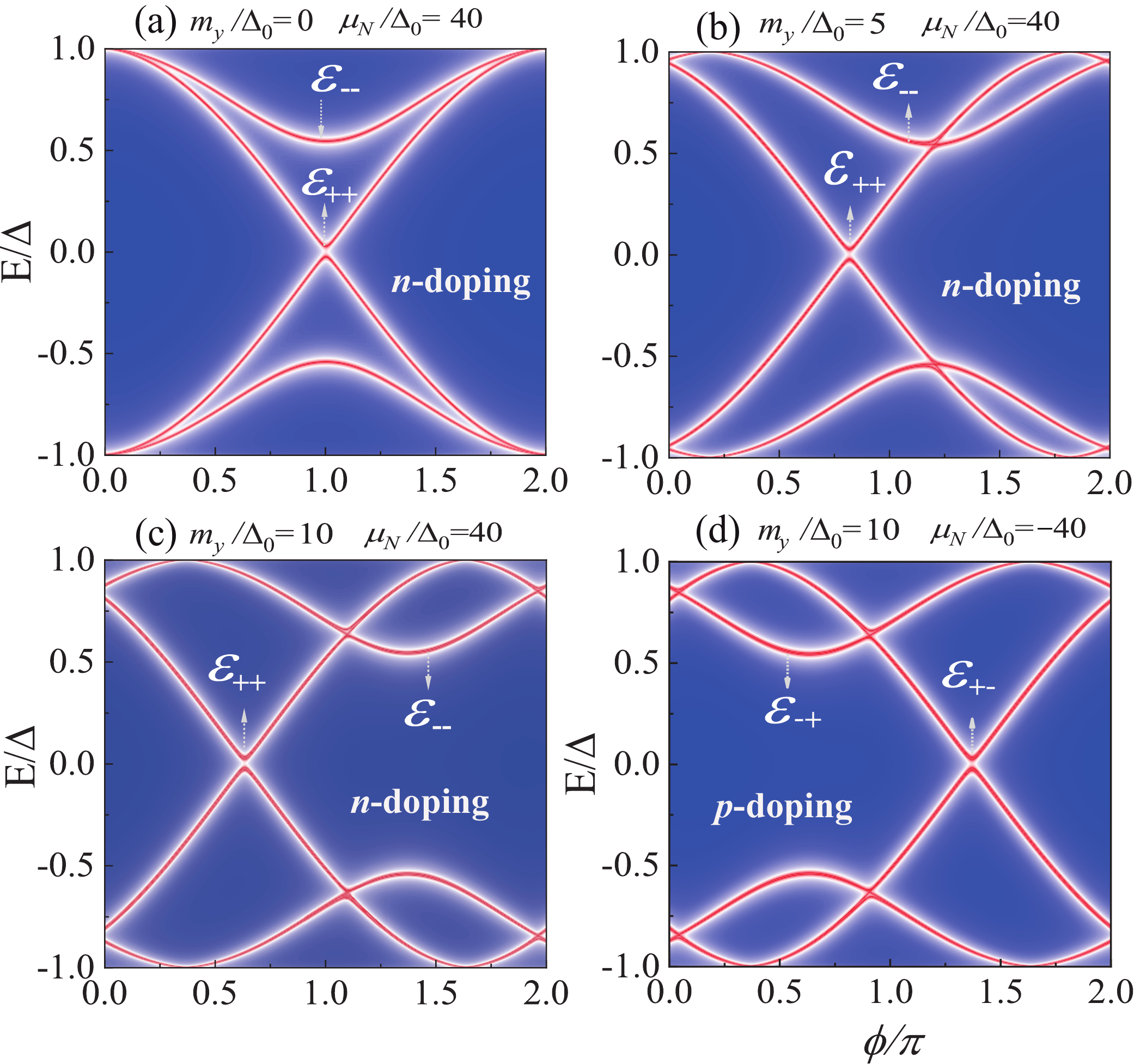}
\caption{Numerically calculated ABS energies as a function of the superconducting phase difference $\phi$ at $k_y=0$, for [(a),(b),(c)] $n$-type ($\mu_N=+40\Delta_0$) and (d) $p$-type ($\mu_N=-40\Delta_0$) doping. Other parameters are as in Fig.~\ref{fig1}. The phase offset of the ABS dispersion reverses sign between the two cases.}
\label{fig3}
\end{figure}

\begin{figure}
\centering
\includegraphics[width=\linewidth]{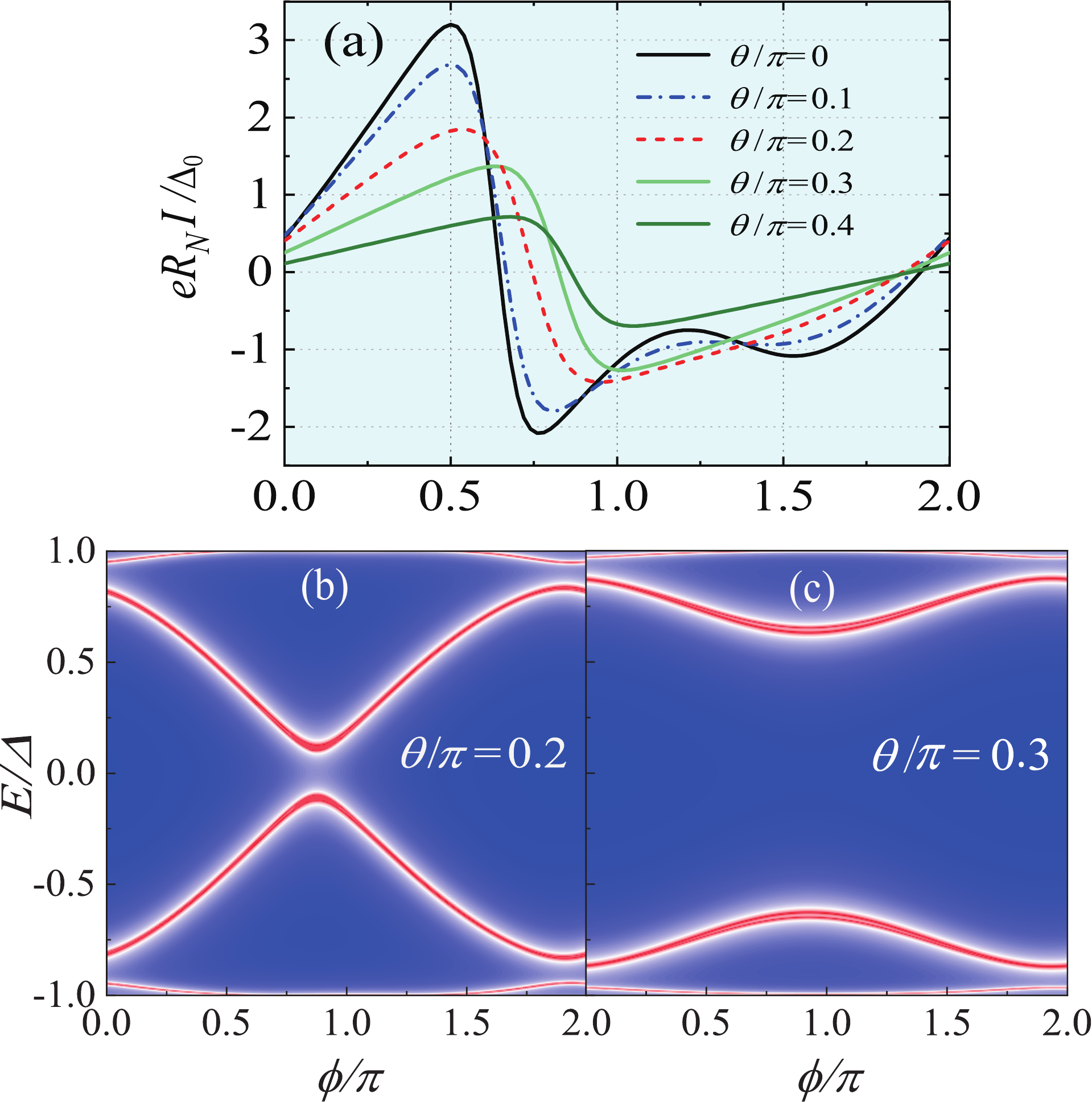}
\caption{(a) $\theta$-resolved CPRs for various incidence angles corresponding to the total CPR in Fig.~1(b). (b), (c) ABS spectra for $\theta=0.2\pi$ and $0.3\pi$, respectively.}
\label{fig5}
\end{figure}

For $n$-type doping, the Fermi level lies in the conduction band, and transport occurs via the bands $(\mu,\nu)=(+,+)$ and $(-,-)$, corresponding to the Fermi surfaces in Fig.~\ref{fig2}(c). Expanding Eq.~(\ref{eq:CPR_from_ABS}) yields
\begin{align}
I_{++} &\simeq a_1 \sin(\phi + \delta\phi_{++}) + a_2 \sin(2\phi + 2\delta\phi_{++}) + \cdots, \\
I_{--} &\simeq \tilde{a}_1 \sin(\phi + \delta\phi_{--}) + \tilde{a}_2 \sin(2\phi + 2\delta\phi_{--}) + \cdots,
\end{align}
with $\delta\phi_{++} = -\delta\phi_{--}$ due to the opposite momentum shifts shown in Fig.~\ref{fig2}(c). The RSOC-induced gap $E_{--}-E_{-+}=6t_R$ causes an amplitude imbalance between $I_{++}$ and $I_{--}$. The total CPR thus becomes
\begin{equation}
I(\phi) \approx I_1 \sin(\phi + \delta\phi_1) + I_2 \sin(2\phi + \delta\phi_2).
\label{eq:total_CPR}
\end{equation}
The relative phase difference $\delta = 2\delta\phi_1 - \delta\phi_2$, controlled by both $m_y$ and the RSOC gap, governs the diode asymmetry.

For $p$-type doping, the Fermi level lies in the valence band, and transport occurs via $(+,-)$ and $(-,+)$, corresponding to the Fermi surfaces in Fig.~\ref{fig2}(d). Since holes have opposite group velocity to electrons, the effective momentum shift in Eq.~(\ref{eq:momentum_shift}) reverses sign relative to the $n$-doped case, leading to $I(\phi) \approx I_1 \sin(\phi - \delta\phi_1) + I_2 \sin(2\phi - \delta\phi_2)$ and hence a reversal of the diode polarity. This gate-controlled sign change of $\delta$ is a key result of our work. Fig.~\ref{fig3} shows the numerically computed ABS spectra for both doping types, clearly demonstrating the sign reversal of the phase offset predicted by Eq.~(\ref{eq:ABS_energy}).

\subsection{Role of transverse momentum channels}

It should be emphasized that the total CPR is the superposition of contributions from all transverse channels. In the wide-junction limit ($W\gg L$), the sum over $k_y$ can be replaced by an integral over the incidence angle $\theta$:
\begin{equation}
\sum_{k_y} \rightarrow \frac{\mu_N W}{2\pi\hbar v_F}\int d\theta\,\cos\theta .
\end{equation}
Figure~\ref{fig5}(a) displays the $\theta$-resolved CPRs corresponding to the total CPR in Fig.~\ref{fig1}(b). At normal incidence ($\theta=0$), the asymmetry between positive and negative branches is most pronounced. As $\theta$ increases, the anharmonicity is strongly suppressed, and the CPR gradually becomes a shifted sinusoid $I(\phi,\theta)\sim \sin(\phi+\phi_0)$ with a reduced phase shift. Thus, both the nonreciprocal critical currents and the $\phi_0$ shift are dominated by small-angle trajectories.

The corresponding ABS spectra in Figs.~\ref{fig5}(b) and (c) confirm this trend. For $\theta=0.2\pi$, the ABS slopes decrease and the phase offset diminishes relative to the $k_y=0$ case, owing to enhanced normal reflection at oblique incidence. At $\theta=0.3\pi$, one of the spin-split ABS branches becomes nearly flat (pinned to the gap edge) and carries negligible supercurrent, leaving only a single shifted ABS that produces no diode effect.

These observations validate our focus on the $k_y=0$ channel for capturing the essential physics of the JDE. Importantly, this graphene mechanism—relying primarily on the normal-incidence channel—is fundamentally distinct from that in conventional 2DEG systems, where the interplay among multiple $k_y$ channels with different $\phi_0$ shifts is essential \cite{Yokoyama2014,Costa2023,Baumgartner2022}.

\begin{figure*}
\centering
\includegraphics[width=0.7\linewidth]{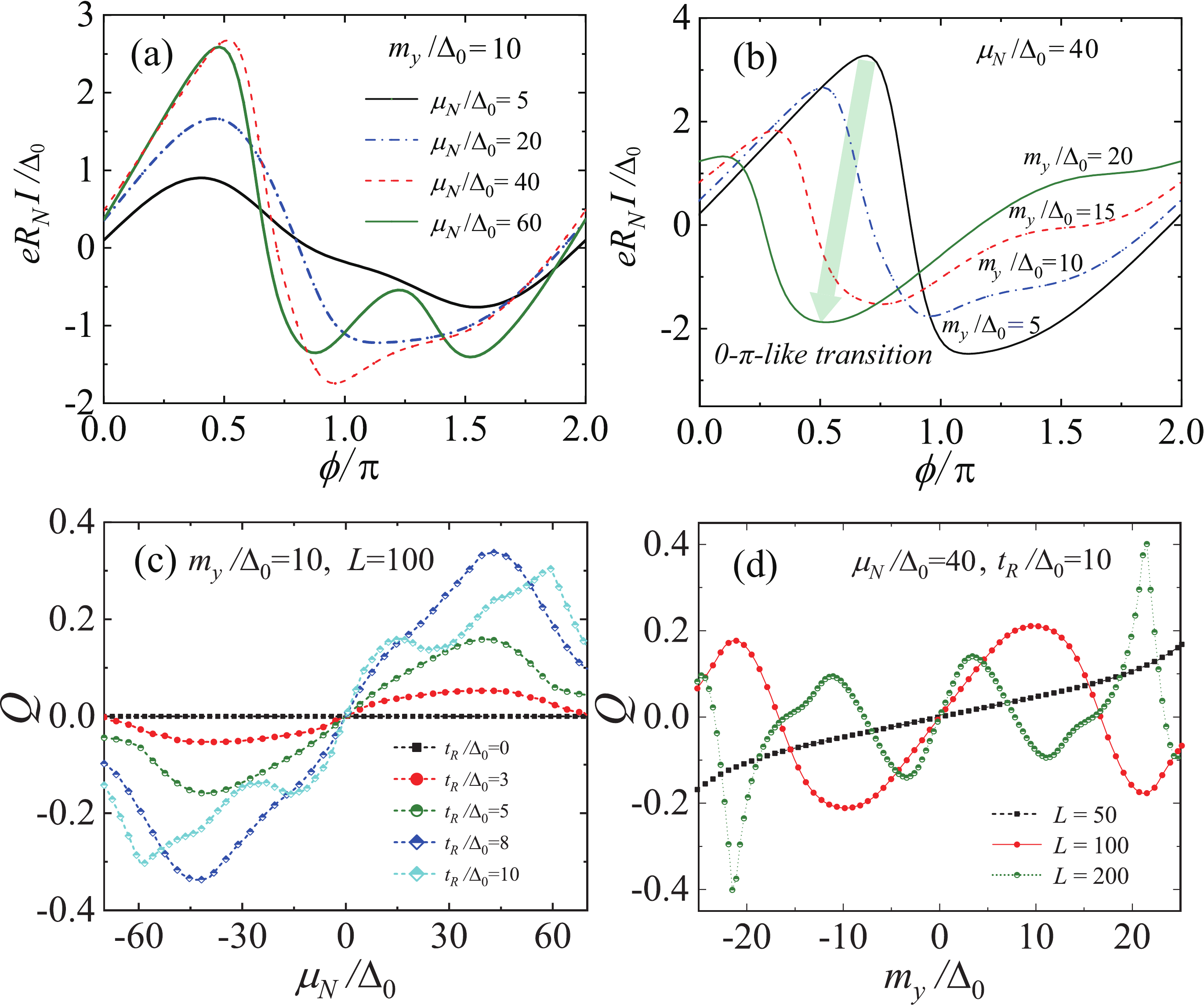}
\caption{(a) CPR for different doping levels $\mu_N$ at fixed $m_y=10\Delta_0$. (b) CPR for different Zeeman fields $m_y$ at fixed $\mu_N=40\Delta_0$. (c) Diode quality factor $Q$ versus $\mu_N$ for several RSOC strengths $t_R$. (d) $Q$ versus $m_y$ for several junction lengths $L$. Other parameters as in Fig.~\ref{fig1}.}
\label{fig4}
\end{figure*}

\section{Evolution of CPR and Diode Efficiency}
\label{sec:evolution}

We now examine how the CPR evolves with the doping level $\mu_N$ and the external Zeeman field $m_y$, and how these parameters control the diode quality factor $Q$. Fig.~\ref{fig4}(a) presents the CPRs for several representative values of $\mu_N$ at fixed $m_y = 10\Delta_0$ and $t_R = 10\Delta_0$. At low doping, e.g., $\mu_N = 5\Delta_0$, the CPR is nearly sinusoidal and exhibits only a small phase shift, resulting in a negligible difference between the positive and negative critical currents. As $|\mu_N|$ increases, the CPR becomes increasingly skewed: the second harmonic component $\sin(2\phi)$ grows significantly relative to the first harmonic, and the overall phase offset becomes more pronounced. This enhanced skewness is the microscopic origin of the diode effect, because a symmetric CPR (purely sinusoidal) cannot produce nonreciprocal critical currents regardless of phase shifts. The emergence of higher harmonics is a direct consequence of the non-parabolic Dirac dispersion and the spin-dependent momentum shifts $\delta q_{\mu\nu}$ discussed in  Sec.~\ref{sec:band}, which break the equivalence between $\phi$ and $-\phi$ at the level of the ABS spectrum.

The doping dependence of the diode efficiency $Q$ is quantified in Fig.~\ref{fig4}(c) for different RSOC strengths $t_R$. At charge neutrality ($\mu_N = 0$, the Dirac point), $Q$ vanishes identically. This is expected because the spin polarization $S_{\mu\nu}(q)$ vanishes as $q \to 0$, and the MCA, which requires a finite spin-momentum locking, is therefore absent. As the Fermi level moves away from the Dirac point, $Q$ grows approximately linearly with $|\mu_N|$ over a wide range, reaching values as high as $\sim 35\%$ for $|\mu_N| \sim 40\Delta_0$ when $t_R = 10\Delta_0$. This linear increase reflects the growing Fermi-surface radius, which enhances the momentum shift $\delta q_{\mu\nu}$ [see Eq.~(\ref{eq:momentum_shift})] and concomitantly increases the phase accumulation asymmetry between electron and hole trajectories. Importantly, for a fixed $t_R$, the efficiency does not increase indefinitely; when $|\mu_N|$ becomes much larger than $t_R$, the RSOC becomes a relatively weak perturbation compared to the kinetic energy, and the spin splitting of the Fermi surfaces saturates. Consequently, $Q$ reaches a maximum at an optimal doping level and then gradually decreases, exhibiting a non-monotonic behavior that is characteristic of the competition between the Dirac linear dispersion and the proximity-induced spin-orbit interaction. This non-monotonicity is markedly different from conventional 2DEG-based Josephson diodes, where $Q$ typically increases monotonically with carrier density because the Rashba energy scales with the Fermi wave vector. In graphene, the chiral nature of the bands introduces an intrinsic upper bound to the spin-orbit effects once the Fermi energy far exceeds the Rashba gap.

Fig.~\ref{fig4}(b) shows the evolution of the CPR with the Zeeman field $m_y$ at fixed $\mu_N = 40\Delta_0$ and $L = 100$. As $m_y$ increases from $5\Delta_0$ to $15\Delta_0$, the CPR not only develops a larger phase shift but also undergoes a progressive deformation, with the position of the maximum supercurrent shifting continuously. At $m_y = 15\Delta_0$, the CPR becomes strongly asymmetric and the ground-state phase difference (the value of $\phi$ where $I = 0$ and the slope is positive) shifts by nearly $\pi$ relative to the $m_y = 0$ case, signaling a $0$-$\pi$-like transition. Such transitions occur when the accumulated phase shift $\delta\phi_{\mu\nu} \simeq 2\delta q_{\mu\nu}L$ crosses odd multiples of $\pi$, effectively reversing the sign of the supercurrent for a given phase bias. The corresponding quality factor $Q$ as a function of $m_y$ is displayed in Fig.~\ref{fig4}(d) for several junction lengths $L$. For short junctions ($L = 50$), $Q$ increases monotonically with $m_y$ in the displayed range, reflecting the linear dependence of $\delta q_{\mu\nu}$ on the Zeeman field. For longer junctions ($L = 100$ and $L = 150$), however, $Q(m_y)$ exhibits pronounced oscillations and even changes sign at larger fields. These sign reversals are directly linked to the $0$-$\pi$-like transitions: each time $\delta\phi_{\mu\nu}$ passes through $(2n+1)\pi$, the diode polarity flips, as seen in the CPR of Fig.~\ref{fig4}(b). The oscillation period is determined by the condition that $\delta\phi_{\mu\nu}$ crosses odd multiples of $\pi$, which depends on the product $m_y L$ and is independent of the RSOC strength to first order, highlighting the geometric nature of the phase accumulation. Notably, these oscillations remain undamped in our clean ballistic model, a feature that arises from the absence of disorder-induced dephasing in graphene's high-mobility channels and distinguishes our predictions from the typically damped oscillations observed in diffusive 2DEG devices.

A crucial advantage of the graphene platform is the ability to switch between electron and hole doping via a single gate voltage. As argued in Sec.~III, the sign of the effective momentum shift $\delta q_{\mu\nu}$ reverses when the Fermi level is tuned from the conduction band to the valence band, because the group velocity of the carriers changes sign while the spin-momentum locking texture remains tied to the band index. This leads to a reversal of the phase offset in the CPR and consequently a reversal of the diode polarity. Indeed, our full numerical calculations confirm that for $p$-type doping (e.g., $\mu_N = -40\Delta_0$), the CPR is mirrored with respect to $\phi$ compared to the $n$-type case, yielding $Q \to -Q$ for the same magnitudes of $m_y$ and $t_R$. This gate-controlled polarity switching is unique to graphene among the currently known Josephson diode platforms, as conventional 2DEGs cannot be continuously tuned across the neutrality point without losing the two-dimensional confinement or introducing significant disorder. Moreover, the continuous tunability of $Q$ from zero to $\sim 40\%$ by simply adjusting the gate voltage, combined with the additional control provided by $m_y$ and $L$, makes the graphene Josephson diode a highly versatile element for future superconducting logic and quantum circuits, where on-chip programmability of both the magnitude and sign of the diode efficiency is highly desirable.

\begin{figure}[H]
\centering
\includegraphics[width=0.8\linewidth]{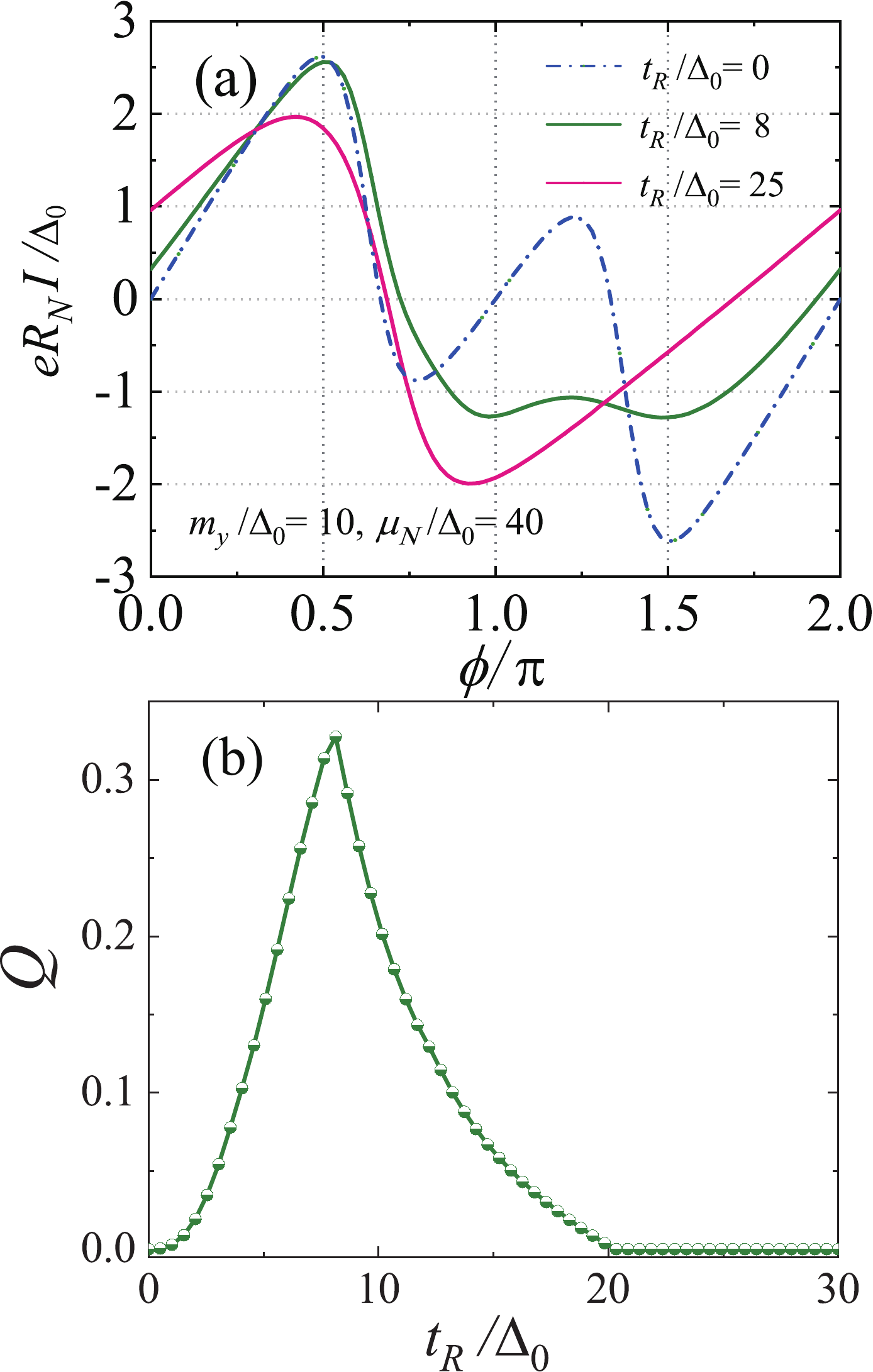}
\caption{(a) Calculated $I(\phi)$ for several values of $t_R$ at $m_y=10\Delta_0$. (b) Diode quality factor $Q$ as a function of $t_R$. Other parameters as in Fig.~\ref{fig1}.}
\label{fig6}
\end{figure}

In addition to the doping and Zeeman field, the Rashba spin–orbit coupling strength $t_R$ also plays a crucial role. In Fig.~\ref{fig6}(a), we plot the CPRs at fixed Zeeman field ($m_y=10\Delta_0$) for several $t_R$ values. In the absence of RSOC ($t_R=0$), the CPR is strongly nonsinusoidal but remains an odd function of $\phi$, so no diode effect is observed. Finite RSOC introduces a cosine term and distorts the CPR symmetry, leading to diode effect. However, for relatively large $t_R$ (e.g., $t_R=25\Delta_0$), the CPR becomes a standard sinusoidal curve with a $\phi_0$-phase shift, implying vanishing of higher harmonics and thus the JDE. The quality factor $Q$ as a function of $t_R$ shown in Fig.~\ref{fig6}(b) displays a nonmonotonic behavior: it initially increases, reaches a maximum, and then decreases. This can be explained by a trade-off: increasing $t_R$ enhances spin-splitting and amplitude difference between spin channels, promoting diode effect, but also acts as a strong scattering barrier that suppresses the Cooper pair tunneling through the gapped channel. When $t_R$ is large enough that Fermi level lies inside the spin-splitting gap, the gapped channel contribution vanishes, and the CPR reduces to a shifted sinusoidal with negligible diode effect.

Additional details, including the free-energy analysis and symmetry considerations, are provided in Appendix~\ref{app:free_energy} and Appendix~\ref{app:symmetry}.

\bigskip

\section{Conclusion}

We have demonstrated that a graphene SNS junction with proximity-induced Rashba spin--orbit coupling and an in-plane Zeeman field functions as a highly gate-tunable Josephson diode. The nonreciprocal critical current not only attains a large efficiency (up to $\sim$40\%) but also reverses its sign when the carrier type is switched from $n$- to $p$-doping via a gate voltage. This gate-controlled polarity switching, together with continuous efficiency modulation from zero to maximum, stems from the chiral nature of Dirac fermions in graphene and has no direct counterpart in conventional 2DEG systems. Furthermore, we predict gate- and field-tunable $0$--$\pi$-like transitions in the current--phase relation, consistent with recent experiments \cite{Costa2023}. Our work establishes graphene as a uniquely versatile platform for superconducting diode phenomena and opens the door to fully gate-programmable Josephson devices for advanced superconducting circuits and quantum technologies.

\section*{Acknowledgments}

This work was supported by the Youth Research Project of Shangrao Normal University (Grant No.~2026SRSYQNKY018 and No.~2026SRSYQNKY020).

\appendix

\section{Free energy and $0$-$\pi$-like transitions}
\label{app:free_energy}

In this Appendix, we provide additional analysis that complements the main text: the phase-dependent free energy and its relation to the $0$-$\pi$-like transitions discussed in Sec.~\ref{sec:evolution}.
The excitation spectrum of the Josephson junction determines the free energy $F$ \cite{Annunziata2011}:
\begin{equation}
\begin{aligned}
F(\phi)&=-k_BT\ln Z\\
       &=-k_BT\ln{\rm Tr}\left[{e^{-\hat{H}_{sns}(\phi)/k_BT}}\right]\\
       &=-k_BT\ln\left[\prod_{n}(1+e^{-\varepsilon_{n}/k_BT})\right]\\
       &=-2k_BT\mathop{\sum}_{n>0}\ln\left[2\cosh(\frac{\varepsilon_{n}}{2k_BT})\right],
\end{aligned}
\end{equation}
where $\varepsilon_{n}$ ($0\leq\varepsilon_{n}\leq\Delta$) are the discrete ABS energies within the superconducting gap. In the short-junction limit, the Josephson current is adequately described by these subgap contributions:
\begin{equation}
I(\phi)=\frac{2e}{\hbar}\frac{\partial F(\phi)}{\partial\phi}
       =-\frac{2e}{\hbar}\sum_{n}\frac{\partial\varepsilon_{n}}{\partial\phi}
        \tanh\left(\frac{\varepsilon_{n}}{2k_BT}\right).
\end{equation}
Conversely, given the CPR $I(\phi)$, the free energy can be obtained as $F(\phi)\propto\int_0^\phi d\varphi\,I(\varphi)$, which allows the determination of the junction's ground state.

\begin{figure}
\centering
\includegraphics[width=0.8\linewidth]{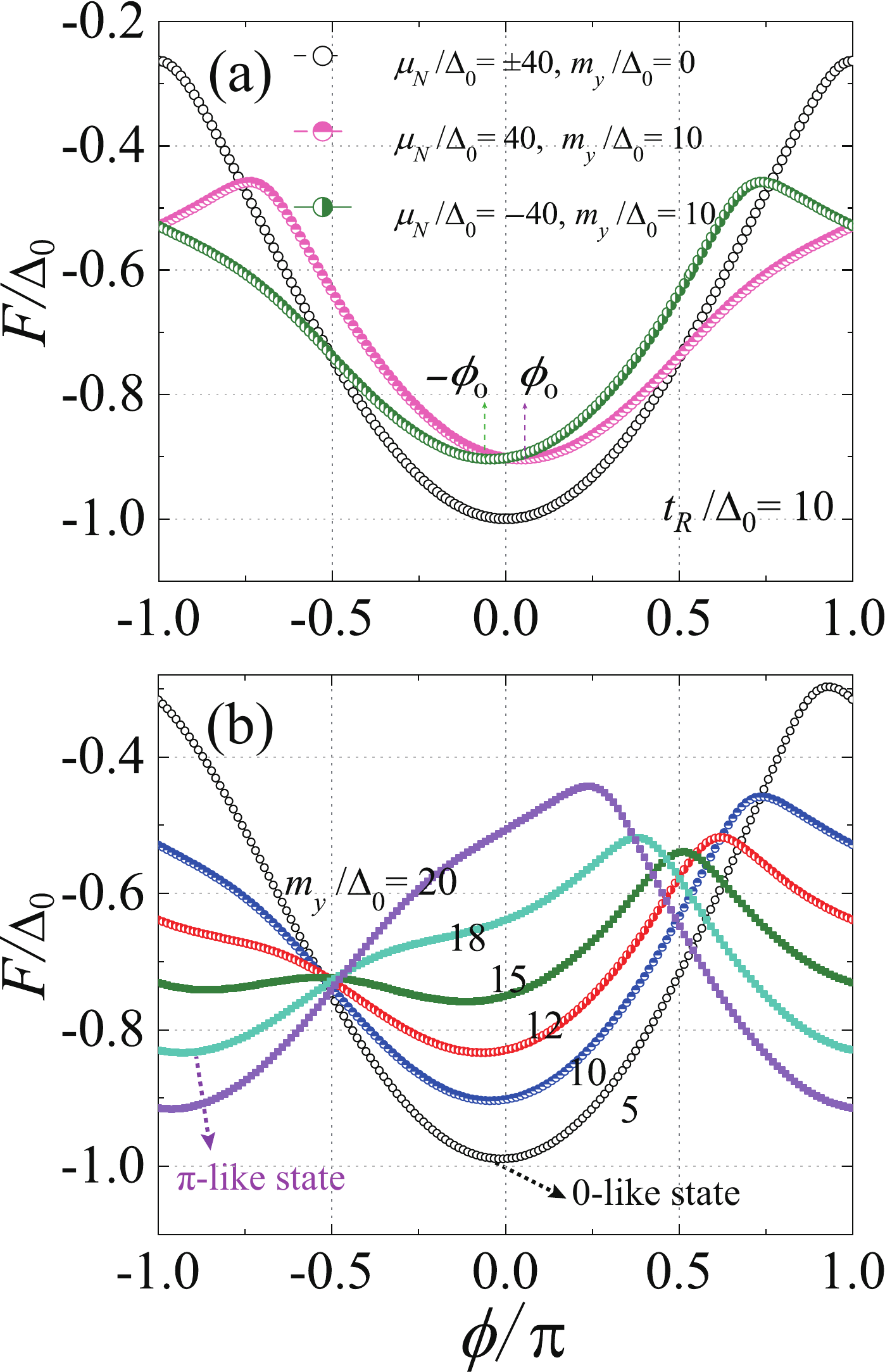}
\caption{(a) Free energy $F(\phi)$ for $\mu_N=\pm 40\Delta_0$ at $m_y=10\Delta_0$. (b) $F(\phi)$ for several $m_y$ values at $\mu_N=40\Delta_0$. Other parameters: $t_R=10\Delta_0$, $L=100$, $T/T_c=0.1$.}
\label{figS3}
\end{figure}

Figure~\ref{figS3}(a) displays $F(\phi)$ for $n$- and $p$-type doping with $m_y=10\Delta_0$ and $t_R=10\Delta_0$. In the absence of Zeeman field, the junction is in a pure $0$-state with minimum at $\phi=0$. The Zeeman field shifts the ground state to $\phi=\phi_0$, yielding a $\phi_0$-junction. Critically, the JDE arises from the asymmetry of $F(\phi)$ about $\phi_0$, not from the $\phi_0$ shift itself. The $n$- and $p$-doped cases exhibit opposite ground-state shifts and opposite asymmetries, corresponding to the reversal of diode polarity discussed in Sec.~\ref{sec:evolution}.

Figure~\ref{figS3}(b) shows $F(\phi)$ for various $m_y$ at fixed $\mu_N=40\Delta_0$. For $m_y<5\Delta_0$, the junction is in a $0$-like state, while for $m_y\geq15\Delta_0$, it enters a $\pi$-like state, signaling a $0$-$\pi$-like transition. These transitions are accompanied by current reversals and sign changes of the diode quality factor $Q$, as observed in Fig.~\ref{fig4}(d). We note that in the absence of RSOC, the current reduces to $I(\phi)\sim I_0\sin(\phi)\cos\delta\phi_{\mu\nu}$, and only a real $0$-$\pi$ transition can occur, as in conventional ferromagnetic Josephson junctions.

\section{Symmetry analysis}
\label{app:symmetry}

To understand the origin of the asymmetric ABSs, we analyze the symmetries of the Bogoliubov--de Gennes (BdG) Hamiltonian. We define the low-energy Dirac Hamiltonian for a single valley ($\eta=\pm1$) as
\begin{equation}
\hat{H}_0^{\eta} = \hbar v_F (\eta q_x \tau_x + q_y \tau_y) \otimes \sigma_0,
\end{equation}
where $\tau_i$ are Pauli matrices in sublattice space and $\eta$ labels the K ($+$) and K$'$ ($-$) valleys. The normal-region Hamiltonian including RSOC is
\begin{equation}
\hat{H}_N^{\eta} = \hat{H}_0^{\eta} + \frac{3}{2}t_R(\eta\tau_x\otimes\sigma_y - \tau_y\otimes\sigma_x).
\end{equation}

In the Nambu basis $\psi\equiv(\psi_u^\uparrow,\psi_u^\downarrow,\psi_v^\downarrow,\psi_v^\uparrow)$, the BdG Hamiltonian is $H_{\rm SNS}(\phi)=\mathcal{H}_L+\mathcal{H}_N+\mathcal{H}_R$, with
\begin{equation}
\mathcal{H}_{L(R)}=
\begin{bmatrix}
\hat{H}_0^{\eta}-\mu_S\hat{I} & \Delta e^{i\phi_{L(R)}}\hat{I} \\
\Delta e^{-i\phi_{L(R)}}\hat{I} & -(\mathcal{T}\hat{H}_0^{\eta}\mathcal{T}^{-1}-\mu_S\hat{I})
\end{bmatrix},
\end{equation}
\begin{equation}
\mathcal{H}_N=
\begin{bmatrix}
\hat{H}_N^{\eta}-\mu_N\hat{I} & 0 \\
0 & -(\mathcal{T}\hat{H}_N^{\eta}\mathcal{T}^{-1}-\mu_N\hat{I})
\end{bmatrix}.
\end{equation}
The time-reversal operator is $\mathcal{T}= i(\eta_x\otimes\sigma_y)\mathcal{K}$, where $\eta_x$ flips the valley index and $\mathcal{K}$ denotes complex conjugation.

Under time reversal, $\hat{H}_N^{\eta}$ transforms as $\mathcal{T}\hat{H}_N^{\eta}\mathcal{T}^{-1} = \hat{H}_N^{-\eta}$, since the RSOC term is invariant under $\mathcal{T}$. In the absence of a Zeeman field, the full BdG Hamiltonian satisfies
\begin{equation}
\mathcal{T} H_{\rm SNS}(\phi) \mathcal{T}^{-1} = H_{\rm SNS}(-\phi),
\end{equation}
which enforces $E_n(\phi)=E_n(-\phi)$. Consequently, the free energy $F(\phi)$ is even and the CPR $I(\phi)$ is odd—a symmetric CPR with no diode effect.

Adding an in-plane Zeeman field $m_y\sigma_y$ breaks time-reversal symmetry. Combined with the inversion-symmetry breaking already present due to RSOC, the condition $E_n(\phi)=E_n(-\phi)$ no longer holds. The resulting CPR becomes asymmetric: $I(\phi)\neq -I(-\phi)$, yielding both the anomalous $\phi_0$ phase shift and the nonreciprocal critical currents $I_c^+\neq I_c^-$ that constitute the Josephson diode effect reported in the main text.


\begin{thebibliography}{99}
\bibitem{footnote}
We assume clean, smooth interfaces and an armchair-type edge along the junction. The unit cell length along $x$ is $\sqrt{3}a/2$.
\end{thebibliography}
\end{document}